\documentclass[thmsa,11pt]{article}
\usepackage{amsfonts}

\begin{document}

\title{\textbf{Naked Singularities, Topological Defects and Brane Couplings}\footnote{To appear in \textit{Quantum Gravity and the Foundations of Physics}, conference held in honor of Prof. Mario Castagnino's 75th birthday, Rosario,
Argentina, March 2010.}}
\author{Jos\'{e} D. Edelstein\,$^{a,b}$, Alan Garbarz\,$^{c}$, Olivera Mi\v{s}kovi\'{c}\,$^{d}$ and Jorge Zanelli\,$^{b}$
\medskip \\
$^{a}${\small \emph{Department of Particle Physics and IGFAE, University of Santiago de Compostela,}}\\
{\small \emph{E-15782 Santiago de Compostela, Spain.}}\\
$^{b}${\small \emph{Centro de Estudios Cient\'{\i}ficos (CECS), Casilla 1469, Valdivia, Chile.}}\\
$^{c}${\small \emph{Departamento de F\'{\i}sica, Universidad de Buenos Aires, Ciudad Universitaria,}}\\
{\small \emph{Pabell\'on 1, 1428, Buenos Aires, Argentina.}}\\
$^{d}${\small \emph{Instituto de F\'{\i}sica, Pontificia Universidad Cat\'olica de Valpara\'{\i}so,}} \\
{\small \emph{Casilla 4059, Valpara\'{\i}so, Chile.}}\medskip \\
{\small jose.edelstein@usc.es, alan@df.uba.ar, olivera.miskovic@ucv.cl, z@cecs.cl}}
\maketitle

\abstract{A conical defect in $2+1$ anti-de Sitter space is a BTZ solution with a negative mass parameter. This is  a naked singularity, but a rather harmless one: it is a point particle. Naturally, the energy density and the spacetime curvature have a $\delta$-like singularity at the apex of the conical defect, but that doesn't give rise to any unphysical situations. Since the conical solution implies the presence of a source, applying reverse enginnering, one can identify the coupling term that is required in the action to account for that source. In that way, a relation is established between the identification operation that gives rise to the topological defect and the interaction term in the action that produces it. This idea has a natural extension to higher dimensions, where instead of a point particle (zero-brane) one finds membranes of even spatial dimensions ($p$-branes, with $p=2n$). The generalization to other abelian and nonabelian gauge theories --including (super-) gravities-- is fairly straightforward: the $2n$-brane couple to a ($2n+1$) Chern-Simons form. The construction suggests a generic role for Chern-Simons forms as the natural way to couple a gauge connection to a brane and avoids the inconsistency that results from the minimal coupling between a brane and a fundamental $p$-form field.}

\section{Introduction}

Naked singularities, like some other forms of nakedness, have an undeserved bad reputation. They allow indiscreet observers to peek into embarrassing secrets of the naked object, inducing to reprovable behavior that violate the rules of moral conduct. Thus, naked singularities are normally required to be hidden behind a geometric veil called horizon. If these were the only drawbacks of naked singularities, it would be a matter of taste whether they should be tolerated or not, as it happens with nudity that is condoned in art, but raises eyebrows and causes scandal in other contexts. This is also the case in physics: a naked singularity might be acceptable in some cases and we call them point particles, cusps, boundaries, branes, or the Big-Bang; in other cases, as in the Schwarzschild solution with negative mass parameter, we consider them as major violations to the laws of physics since, as J. Earman pointed out, ``Green slime, lost socks and TV sets could emerge from them" \cite{Earman}. It has been argued that nature must abhor these singular objects and some higher principle --aptly called cosmic censorship--, is often invoked to avoid them \cite{Penrose-censorship}.

The laws of physics, however, do not depend on opinion and the banning of naked singularities should be carefully scrutinized, as we will see it is the case for extended objects such as branes. In the past, the coupling between branes and fundamental gauge fields in a form that is a direct extension of the minimal coupling of the form $j^{\mu_1 \cdots \mu_p} A_{\mu_1 \cdots \mu_p}$, was shown to lead to inconsistencies in the Hamiltonian formalism in the case of nonabelian theories \cite{Teitelboim:1985ya}. As we will see next, branes coupled to Chern-Simons forms --Chern-Simons branes--, are a class of naked singularities that are extended sources for gauge theories and for gauge (super-)gravities, and not necessarily sources of paradoxes.

Mario Castagnino was an inspiring force behind theoretical physics during obscure times in Argentina and the neighboring countries. He encouraged young people to follow their own wild ideas and to pay no heed to any form of censorship, be it cosmic or otherwise. It is therefore fitting on this occasion to dedicate to Mario this essay on such a class of naked singularities.

\section{Remarks on the BTZ spectrum}

Gravity in $2+1$ dimensions is a remarkably simple and useful model that mimics some aspects of our real ($3+1$) universe. A particularly simple feature of $2+1$ AdS gravity is the existence of a black hole, the so-called BTZ solution \cite{BTZ},
\begin{equation}
ds^2=-f^2 dt^2 + f^{-2}dr^2 + r^2 (d\phi - N dt)^2 ~,
\end{equation}
where $f^2=-M+r^2/\ell^2+J^2/4r^2$ and $N=J/2r^2$. This metric describes a black hole of mass $M$, and angular momentum $J$, and two horizons located
at $r=r_{\pm }$, where \footnote{Note that in 2+1 dimensions the gravitational constant has dimensions of length and the cosmological constant in AdS spacetime is $\Lambda =-\ell^{-2}$, so that the mass parameter $M$ is dimensionless in geometric units.}
\begin{equation}
r^2_{\pm }=\frac{M\ell^2}{2}\left( 1\pm \sqrt{1-\frac{J^2}{M^2\ell^2}}
\right) ~,
\end{equation}
which are clearly positive provided $M\ell > |J|$ and they coincide if $M\ell = |J|$. On the other hand, for $M\ell < |J|$ the spacetime exhibits a naked conical singularity, except for the isolated point $(J, M)=(0,-1)$ which is the globally AdS spacetime. This supported the view adopted in Ref.\cite{BTZ} according to which the physical spectrum of the 2+1 black hole included the massive states
$M\ell \geq |J|>0$, the massless black hole, $M=0$, and the ``ground state'' represented by AdS space, separated from the other physical states by a gap of forbidden naked singularities. The black hole is also a solution of the supersymmetric extension of 2+1 AdS gravity,
and it was observed that for some values of $M$ and $J$ these states are invariant under supersymmetry, also called BPS states. Following the standard reasoning of Olive and Witten \cite{Olive-Witten}, one can prove that these states must be stable under perturbative deformations and they are suitable ground states for perturbative quantization. These special cases are the extremal ($M\ell = |J|$), the massless ($M=0$) black holes, and the AdS vacuum ($M=-1$) \cite{Coussaert-Henneaux}.

The 2+1 black hole has an obscure side, namely, the part of the spectrum with $0>M>-1$. As noted in \cite{Izquierdo-Townsend}, those are naked singularities obtained by identifications with a Killing vector that rotates the spatial section $\Sigma =\{t=0\}$ about a fixed point, $r=0$, in AdS. As mentioned above, naked singularities can be unphysical sources of paradoxes, but a conical singularity need not be a challenge to reality. It is at most an idealized situation where a very large curvature is concentrated on a very small area, something that in theoretical physics is described by a Dirac delta in a standard form. It was also shown that these solutions, when endowed with angular momentum behave very much like their $M>0$ black hole relatives \cite{MZ-1}. Moreover, if their angular momenta matches their masses, these spinning conical singularities are also BPS states.

\section{Source for a conical defect}

Identifying $x \sim \exp(2\pi (1 - \alpha)\,\mathbf{J}_{12})\, x$ is equivalent to removing a wedge of size $2\pi \alpha$ and pasting along the edges. The result is a manifold of negative constant curvature everywhere, except for a line of infinite curvature at $r = \sqrt{(x^1)^2 + (x^2)^2} = 0$. This is the locus of the naked singularity where the spatial section has a conical defect of magnitude $2\pi\alpha$.

But, as noted long ago in \cite{Deser-Jackiw-tHooft}, there is a problem here: the new geometry is not a solution of the homogeneous Einstein field equations. The conical defect produces a $\delta $-like curvature singularity, which must appear on the right hand side of the equations,
\begin{equation}
R^{ab}+\frac{1}{\ell ^{2}}\,e^{a}\wedge e^{b}=2\pi \alpha \,\delta _{\Gamma }(\Sigma )\,\delta _{12}^{ab}~,
\label{Einst-eq}
\end{equation}
where $\delta _{\Gamma }(\Sigma )=\delta (x^{1},x^{2})\;dx^{1}\wedge dx^{2}$ is the two-form Dirac delta integrated over the the spatial section $\Sigma$ (the $x^{1}$-$x^{2}$ plane) supported on the $(D-2)$-dimensional worldvolume $\Gamma $,
\footnote{Support on $\Gamma $ for the Dirac $\delta $-function $\delta _{\Gamma }(\Sigma )$ means that it restricts the dynamics of the $p$-forms $f(x)$ to the submanifold $\Gamma $, that is, $\int_{\Gamma \times \Sigma }f(x)\wedge \delta _{\Gamma }(\Sigma )=\int_{\Gamma }f(x)$.} and $R^{ab}=(d\omega +\omega \wedge \omega )^{ab}$ is the curvature 2-form. We assume that torsion vanishes for the sake of simplicity, $T^{a}=0$. This in turn means that the equations are not obtained varying the free Einstein-Hilbert
action, but there is a source term that must couple to the geometry to account for the extra term in the equations. The form of this coupling term is best described in the first order representation, where the vielbein ($e^{a}$) and the Lorentz connection for $SO(1,2)$ ($\omega ^{ab}$) are combined into a connection for $SO(2,2)$,
\begin{equation}
\mathbf{A}=\frac{1}{2}\,\omega^{ab}\mathbf{J}_{ab}+\frac{e^{a}}{\ell }\, \mathbf{J}_{a}\,,\quad \mbox{and}\quad a,b\in
\left\{ 0,1,2\right\} ~,
\end{equation}
where $\mathbf{J}_{ab}$ and $\mathbf{J}_{a}$ generate the AdS algebra in 2+1 dimensions $so(2,2)$. One advantage of this compact notation is that the three-dimensional gravitational Lagrangian, $L\equiv \sqrt{|g|}(R-2\Lambda )$ can be written as a Chern-Simons (\textbf{CS}) form,\footnote{See, for instance Ref. \cite{Z-BsAs}. In what follows, wedge product of forms will be omitted, except when explicitly needed for clarity.}
\begin{eqnarray}
L\,d^{3}x &=&\epsilon _{abc}\left( R^{ab}\wedge e^{c}+\frac{1}{3\ell ^{2}}\,e^{a}\wedge e^{b}\wedge e^{c}\right)   \nonumber \\
&=&\left\langle \frac{1}{2}\,\mathbf{A}\wedge d\mathbf{A}+\frac{1}{3}\, \mathbf{A}\wedge \mathbf{A}\wedge \mathbf{A}\right\rangle ~\equiv ~\left\langle \mathcal{C}_{3}(\mathbf{A})\right\rangle ~,
\end{eqnarray}
where $\left\langle \cdots \right\rangle $ is an invariant trace in the algebra, defined as
\begin{equation}
\left\langle \mathbf{J}_{AB}\,\mathbf{J}_{CD}\right\rangle =\epsilon_{ABCD}~,\quad \mbox{and}\quad A,B,\ldots \in
\left\{ 0,1,2,3\right\} .
\end{equation}

The action that yields (\ref{Einst-eq}), together with $T^a = 0$, is found to be
\begin{equation}  \label{Action+J}
I[\mathbf{A};\mathbf{j}]=\int \left\langle \mathcal{C}_3(\mathbf{A}) - \mathbf{j} \wedge \mathbf{A}\right\rangle ~,
\end{equation}
where the current is given by
\begin{equation}  \label{current}
\mathbf{j}=2\pi\alpha \,\delta_\Gamma(\Sigma)\, \mathbf{J}_{12} ~,
\end{equation}
and the field equations (\ref{Einst-eq}) and $T^a = 0$ take the simple form
\begin{equation}
\mathbf{F}=\mathbf{j}\,.
\end{equation}
Note that this current is such that it takes the projection of the connection along the worldline of the charge, $e^0$. The current has the
general form
\begin{equation}  \label{current-defect}
\mathbf{j}=\left(
\begin{array}{c}
\mbox{Magnitude} \\
\mbox{of defect}
\end{array}
\right) \times \left( \delta \left[
\begin{array}{c}
\mbox{supported on the } \\
\mbox{worldline }\Gamma
\end{array}
\right] \right) \times \left(
\begin{array}{c}
\mbox{Generator of} \\
\mbox{identification } \mathbf{J}
\end{array}
\right) ~,
\end{equation}
which will be taken as the prototype for the coupling between branes and connections. In the case (\ref{current}), it is a 0-brane minimally coupled to the nonabelian connection 1-form for the gravitational field, $\mathbf{A}$. This brane is ``charged'' with respect to the gauge field, the charge being the magnitude of the deficit, $2\pi\alpha$, which in the gravitational case is the rest mass of the particle. The following pages will be devoted to the generalization of this idea to higher-dimensional cases and to other connections as well.

\section{Beyond $0$-branes, beyond gravity}

\subsection{Gauge invariance}

A topological defect, as any external source, breaks translational invariance (invariance under AdS boosts in anti de Sitter spaces). An
immediate consequence of the coupling is the breaking of the original $SO(2,2)$ gauge symmetry to the subgroup spanned by those generators that commute with $\mathbf{J}_{12}$, namely, the invariance group of the worldline of a particle at rest, $SO(2)\times \mathbb{R}$. This remains true if the conical defect has angular momentum, although in that case the identification is produced by a Killing vector with a different generator (see, e.g., Ref. \cite{MZ-1,EGMZup}).

In view of this broken symmetry of the solution, one concern would be whether this coupling is compatible with the gauge invariance of the theory. It is obvious that the action (\ref{Action+J}) is invariant under a simultaneous gauge rotations of $\mathbf{A}$ and $\mathbf{j}$. However, one needs to ensure that $\mathbf{j}$ does transform in the right representation under the action of the group, and that this transformation is produced dynamically.

Written in the form (\ref{Action+J}), the current $\mathbf{j}$ is an external, nondynamical source of a given strength and sitting at a certain fixed location in space. Now, under a gauge transformation, $\mathbf{A}$ changes by the covariant derivative of an arbitrary Lie-algebra valued zero-form $\mathbf{\Omega }$,
\begin{equation}
\delta \mathbf{A} = D \mathbf{\Omega } = d\mathbf{\Omega }+[\mathbf{A}, \mathbf{\Omega }] ~,
\end{equation}
and hence, the coupling term changes as
\begin{equation}
\delta \left\langle \mathbf{j}\mathbf{A}\right\rangle = \left\langle D(\mathbf{j}\mathbf{\Omega })\right\rangle - \left\langle (D\mathbf{j})\mathbf{\Omega }\right\rangle.
\end{equation}
The first term is a total derivative ($= d \left\langle \mathbf{j}\mathbf{\Omega }\right\rangle$) and thus it changes the action by a surface term, without affecting the field equations. This term can be disposed of under reasonable boundary conditions. The vanishing of the second term --charge conservation-- is a requisite common to any physically meaningful conserved current in a gauge theory,
\begin{equation}
D\mathbf{j}=0 ~,
\end{equation}
where $D\mathbf{j}=d \mathbf{j} +[\mathbf{A},\mathbf{j}]$. It can be checked that the current (\ref{current}) satisfies this
condition: outside $\Gamma$, $\mathbf{j}$ and $d\mathbf{j}$ vanish identically; on $\Gamma$, $d\mathbf{j}=0$ and $[\mathbf{j}, \mathbf{A}]=0$ because $\mathbf{j}$ and $\mathbf{A}$ belong to commuting sectors of the gauge algebra.

A natural way to ensure the gauge invariance is to assume that the current is produced by dynamical matter that belongs to some (vector) representation of the gauge group and is described by a gauge invariant action. Then, as it happens with the quark current in QCD, the current $\mathbf{j}_{\mu}^a=\bar{\psi}\gamma_{\mu} \mathbf{T}^a\psi$ is covariantly conserved by construction
because it is Noether's current associated to the $SU(3)$ invariance of the quark Lagrangian.

\subsection{$(D-3)$-branes}

The above construction can be extended from $3$ to $D=2n+1$ dimensions and, instead of a $0$-brane, one could consider a $2k$-brane, with $0\leq k < n$ \cite{MZ-2}. Among these, the $(D-3)$-brane in $D=2n+1$ dimensions is a particularly simple case, because this brane is the result of an identification by a Killing vector, and is therefore described by locally flat connection (except on sets of measure zero).

Consider a $D$-dimensional globally anti-de Sitter spacetime (AdS$_D$) and a two-dimensional spacelike plane $\Sigma$ in it, with coordinates $x^1$ and $x^2$. This plane is left invariant under the action of a rotation generated by $\mathbf{J}_{12}$. An identification by the same Killing vector is
\begin{equation}
\left(
\begin{array}{c}
x^1 \\
x^2
\end{array}
\right) \sim e^{2\pi (1-\alpha) \mathbf{J}_{12}} \left(
\begin{array}{c}
x^1 \\
x^2
\end{array}
\right),
\end{equation}
and it gives rise to an angular deficit $2\pi \alpha$ on the set $\Gamma=\left\lbrace x^{\mu} | x^1=x^2=0\right\rbrace $, at the center of $\Sigma$, the set of fixed points of the Killing vector. This submanifold of codimension two is the world history of a $(D-3)$-brane. Since the identification is produced by an isometry, the local geometry is not affected and the spacetime is still locally AdS. Therefore, the field equations are those that have as solution AdS$_D$, everywhere except at $\Gamma$, where one should find a $\delta$-like singularity. The field equations read
\begin{equation}
\mathbf{F} = \mathbf{j} = 2\pi \alpha\, \delta_\Gamma(\Sigma)\, \mathbf{J}_{12} \,,  \label{F=j}
\end{equation}
where in the right hand side is the $2$-form $\delta_\Gamma(\Sigma)=\delta(x^1, x^2)\,dx^1\wedge dx^2$.

What action gives rise to this equation? The answer depends on what action one assumes to be the one that gives global AdS$_{D}$ as a solution. In fact, there are many. For instance, any action whose field equations are of the form
\begin{equation}
\epsilon _{a_{1}\cdots a_{D}}\left( R^{a_{1}a_{2}}+\frac{e^{a_{1}}e^{a_{2}}}{\ell^{2}} \right) \cdots \left( R^{a_{2s-1}a_{2s}}+\frac{e^{a_{2s-1}} e^{a_{2s}}}{\ell^{2}} \right) e^{a_{2s+1}}\cdots e^{a_{D-1}}=0,
\end{equation}
with $1\leq s \leq(D-1)/2=n$, would do \cite{Crisostomo-Troncoso-Zanelli}.  The case with the largest possible brane dimension ($k=n-1$) is particularly interesting. In that case, which is only possible in odd dimensions, the bulk dynamics of the geometry is governed by the CS action, and the field equations in the complement of $\Gamma $, take the particularly simple form $\mathbf{F}^{n-1}=0$. Thus, a natural way to produce an equation compatible with (\ref{F=j}) is to couple the source $\mathbf{j}$ to the CS form defined on $\Gamma $ \cite{MZ-1,MZ-2,EGMZ}. In this way, the action is
\begin{equation}
I=\kappa \int\limits_{\mathbb{M}}\left\langle \mathcal{C}_{2n+1}(\mathbf{A})- \mathbf{j}\wedge \mathcal{C}_{2n-1}(\mathbf{A})\right\rangle \,.
\end{equation}
Here, $\mathbb{M}$ is a $D$-dimensional manifold and the level $\kappa $ is a dimensionless quantized coupling constant. The CS density $\mathcal{C}_{2k+1}(\mathbf{A})$ is a ($2k+1$)-form that is polynomial in the connection 1-form $\mathbf{A}$ and the curvature 2-form $\mathbf{F}=d\mathbf{A}+\mathbf{A}\wedge \mathbf{A}$, and is defined through the relation
\begin{equation}
d\langle \mathcal{C}_{2k+1}(\mathbf{A})\rangle =\frac{1}{k+1}\,\langle \mathbf{F}^{k+1}\rangle \,,
\label{CS density}
\end{equation}
and the field equation reads
\begin{equation}
\mathbf{F}^{n-1}\wedge (\mathbf{F}-\mathbf{j})=0\,.  \label{F^n=jF^n-1}
\end{equation}
Note that all solutions of (\ref{F=j}) also solve (\ref{F^n=jF^n-1}), but there are many more solutions in which only the prefactor $F^{n-1}$ vanishes. These extra solutions arise from the degeneracy and irregular character of the Chern-Simons systems in dimensions greater than three \cite{Miskovic-Zanelli-irregular,MTZ,LecturesZ}.

\subsection{Branes of various sizes}

Chern-Simons branes of all dimensions ($0 \leq 2k<D$) can also be coupled in a similar way, via the interaction term \cite{Z-BsAs}
\begin{equation}  \label{general CS coupling}
\int\limits_{\mathbb{M}}\left\langle \mathbf{j}_{[2k]} \wedge \mathcal{C}_{2k+1}(\mathbf{A})\right\rangle ,
\end{equation}
where $\mathbf{j}$ is a ($D-2k-1$)-form supported on the ($2k+1$)-dimensional worldvolume $\Gamma^{2k+1}$ of the brane, and $\mathcal{C}_{2k+1}$ is the CS ($2k+1$)-form living on the brane history.

As mentioned above, a brane that is produced by a single identification with a Killing vector --as is the case if $\Gamma $ has codimension two--, has no effect on the local geometry. This continues to be true if the brane is the result of several simultaneous identifications produced by commuting Killing fields. For instance, in $D=5$, an identification on the planes $(x^{1}$-$x^{2})$ and $(x^{3}$-$x^{4})$ produces a pair of intersecting orthogonal $2$-branes in a locally trivial geometry. As in the case of a single brane, the perturbative stability of these configurations can be ensured if the geometry admits a globally defined Killing spinor, which one expects to be possible if the topological defects carry an abelian charge to make them extremal \cite{EGMZ}. One may also expect that a Killing spinor can be defined provided the defects are also endowed with angular momenta on the two orthogonal planes \cite{EGMZup}.

Configurations with several parallel topological defects produced by removing several wedges are also possible and they generically wouldn't alter the local geometry except at the defects. There seems to be a constraint on the maximum total angular deficit produced by all of them together. The dynamical interaction of these defects is something we expect to report elsewhere.

Note that in the preceding remarks there is no need to restrict to a particular form of the free gravitational action that defines the geometry on the complement of $\Gamma$. In fact, that discussion applies equally for any bulk gravitational action: Einstein-Hilbert with or without cosmological constant, Lovelock gravities, Chern-Simons gravities, etc. An interesting case is the $0$-brane coupled to the CS action in $D=2n+1$ gravity,
\begin{equation}
I[\mathbf{A};\mathbf{j}_{[0]}]=\int\limits_{\mathbb{M}^{2n+1}} \left\langle \mathcal{C}_{2n+1}(\mathbf{A})-\mathbf{j}_{[0]} \wedge \mathbf{A} \right\rangle ~.
\end{equation}
In that case, the naked singularity is a topological defect generated by removing a solid angle in a $\mathbb{S}^{2n-1}$ sphere, that produces a surface deficit. The metric corresponding to this solution is that of a spherical black hole but with the ``wrong sign of mass'' $M<0$ \cite{MZ-1}. This geometry is not of constant curvature, and the curvature is unbounded in the vicinity of the naked singularity. Based on the evidence of analogous situations in general relativity \cite{Dotti-Gleiser-Pullin}, one would expect this geometry to be unstable under perturbations.

BPS branes are easily constructed for $k=(D-3)/2$ since they do not change the local geometry and therefore, it is sufficient to add an abelian gauge connection in a simple configuration --or spin-- to compensate for the nontrivial holonomies due to the angular defect in the geometry, as shown in \cite{MZ-1,EGMZ}. If a brane deforms the local geometry, it would be necessary a fine matching between the local behavior of the compensating field and the geometry to allow for a covariantly constant (\textit{i.e.}, Killing) spinor. It is an interesting question whether this can be achieved.

\subsection{Other groups, general recipe}

The idea of a gauge invariant coupling using the CS form as a generalization of the coupling to a $0$-form is not exclusive of gravitation. The general form (\ref{current-defect}) is equally applicable to other gauge groups and to branes of any dimension, so that instead of (\ref{current}) we can write
\begin{equation}  \label{current-general}
\mathbf{j}_{[2k]}=2\pi\alpha \,\delta_\Gamma(\Sigma)\, \mathbf{G},
\end{equation}
where $\delta_\Gamma(\Sigma)$ is the ($D-2k-1$)-form supported at the center of the transverse space, and \textbf{G} is a multilineal homogeneous function of the generators characterizing the charge of the source. Gauge invariance is partially broken by the presence of an externally fixed, nondynamical source. But again, the invariance is restored if the current is covariantly conserved, which is automatically satisfied if it is produced by dynamical matter fields.

One is thus led to postulate the coupling between a $2k$-brane and a connection $\mathbf{A}$ in any dimension that generalizes the usual minimal coupling of electrodynamics\cite{Z-BsAs},
\begin{equation}
I_{\mathrm{int}}[\mathbf{A};\mathbf{j}_{[2k]}]=\int\limits_{\mathbb{M}^D} \left\langle \mathbf{j}_{[2k]} \wedge \mathcal{C}_{2k+1}(\mathbf{A})\right\rangle .
\end{equation}
This idea was successfully exploited by these authors to CS gravities in dimensions $D \geq 5$ \cite{EGMZ}, where there is the added advantage that the supersymmetric extension is exactly known \cite{Banados:1996hi,Troncoso:1997va}, the corresponding supersymmetric algebra is realized as a gauge supersymmetry \cite{MTZ}, and the BPS bound can be computed explicitly \cite{Miskovic:2006ei}. The idea of coupling extended objects using CS forms was also explored in the context of supergravity \cite{Mora-Nishino}.

\section{Concluding remarks}
\subsection{The role of CS forms in physics}

It is a curious feature of the standard minimal coupling that it involves the gauge potential $\mathbf{A}$, which is not gauge invariant, but \textit{quasi}-invariant: under a gauge transformation it changes by an exact form, $d\mathbf{\Omega }$. In retrospect, this is not surprising, since all CS forms transform in this way under an infinitesimal gauge transformation,
\begin{equation}
\delta \left\langle \mathcal{C}_{2k+1}(\mathbf{A}) \right\rangle = d\mbox{(something)}.
\end{equation}
In electrodynamics, this feature has two consequences besides providing a way to couple the electromagnetic field to charged matter: \emph{a}) it ensures the invariance of the holonomies,
\begin{equation}
\int\limits_{\Gamma} A\,,
\end{equation}
under gauge transformations for closed paths $\Gamma$, and \emph{b}) it requires the conservation of charge,
\begin{equation}
\delta \int j^{\mu} A_{\mu} \,d^Dx = \int j^{\mu} \delta A_{\mu} \,d^Dx=\int j^{\mu} \partial _{\mu} \Omega \,d^Dx = -\int \partial _{\mu} j^{\mu} \Omega \,d^Dx,
\end{equation}
which vanishes for arbitrary $\Omega$ if and only if $\partial_{\mu} j^{\mu}=0$. These are essentially the same features that characterize the CS coupling between an arbitrary nonabelian gauge field and a brane discussed above. Moreover, it seems that CS forms are quite unique in performing these tricks. One might turn this argument around and conjecture that ``the role'' of CS forms in physics is to provide gauge invariant couplings to extended forms of matter.

\subsection{Other gauge theories}

Although the CS form seems to provide a privileged form of coupling, they are clearly not unique as Lagrangian functionals for theories with som local symmetries, as exemplified by the very existence of the Yang-Mills and the Einstein-Hilbert actions. Of course, as emphasized elsewhere \cite{EZsort}, the CS actions for gauge theories also enjoy some outstanding features that make them interesting candidates for fundamental descriptions of nonabelian gauge theories, including gravity.

The most appealing feature of CS gravity is that it is a \textit{bona fide} gauge theory, constructed using only a connection in which the Lorentz (``spin'') connection and the vielbein are combined. As a consequence, the metric does not appear as a fundamental field in the action principle, but as a derived entity in the solutions. In this way, the gravitational field is described by more degrees of freedom than those encoded in the metric alone, which is viewed as a composite emerging feature of the full geometrical content of spacetime \cite{Edelstein-Zanelli}.

In spite of their many attractive features, like having naturally defined supersymmetric extensions \cite{EZsort}, and having no arbitrary dimensional constants \cite{Zanelli:1994ti}, CS theories have the obvious drawback that they only make sense in odd dimensions. For this reason, CS theories are often regarded as toy models with no connection to the real world. That is, however, a rather narrow view because there are interesting physical systems in which the relevant dimension is not necessarily that of ordinary spacetime as, for instance, in superconductivity, whose dynamics is aptly described by a 3-dimensional CS action.

\subsection{Even-dimensional defects}

The coupling mechanism described here only works for $p$-branes with even $p$. But, of course, the cases with odd $p$ are also conceivable. How can the latter case be understood? Is it possible to define an odd-dimensional topological defect (string) and how would it couple to the surrounding spacetime geometry? Obviously they do not couple to CS forms, but there seems to be a different role for these defects. In \cite{Anabalon et al}, it was shown that if an identification by a rotational Killing vector in a six-dimensional spacetime of trivial geometry (of constant negative curvature, AdS) is performed, the 4-dimensional topological defect that is produced resembles our spacetime. More precisely, if the 6D spacetime is governed by a trivial Lagrangian (the 6D Euler form), then the geometry of the topological defect obeys Einstein-Hilbert dynamics.

Naked singularities, topological defects and branes of the sort discussed in this essay have a broad implication in more general scenarios than those provided by Chern-Simons theories. Following the arrow of time, thoroughly studied by Castagnino, we expect that the role that these theories and their extended objects have to play in our understanding of physical systems be uncovered.

\section*{Acknowledgement}

AG and JZ wish to warmly thank the organizers of the meeting \textit{Quantum gravity and the foundations of physics} held in honor of Prof. Mario Castagnino's 75th birthday for the opportunity to be part of a deserved homage. We would like to thank Gast\'{o}n Giribet for interesting discussions. This work is supported in part by MICINN and FEDER (grant FPA2008-01838), by Xunta de Galicia (Conseller\'{\i}a de Educaci\'on), by the Spanish Consolider-Ingenio 2010 Programme CPAN (CSD2007-00042), by a bilateral agreement MINCyT Argentina (ES/08/02) -- MICINN Spain (FPA2008-05138-E), and by FONDECYT Grants 11070146, 1100755 and 7080201.
A.G. was also partially supported by UBA-Doctoral Fellowship 572/08. O.M. is supported by the PUCV through the projects 123.797/2007, 123.705/2010 and MECESUP UCV0602. The Centro de Estudios Cient\'{\i}ficos (CECS) is funded by the Chilean Government through the Millennium Science Initiative and the Centers of Excellence Base Financing Program of Conicyt, and by the Conicyt grant ``Southern Theoretical Physics Laboratory'' ACT-91. CECS is also supported by a group of private companies which at present includes Antofagasta Minerals, Arauco, Empresas CMPC, Indura, Naviera Ultragas and Telef\'onica del Sur.

\end{document}